\documentclass[12pt]{article}
\usepackage{etex}
\usepackage{hyperref}
\hypersetup{
    colorlinks = true,
    linkcolor = blue,
    urlcolor = blue,
    citecolor = blue
}
\usepackage{fontenc}
\usepackage[latin1]{inputenc}
\usepackage{amsmath}
\usepackage{amsfonts}
\usepackage{amssymb}
\usepackage[affil-it]{authblk}
\usepackage{amsthm}
\usepackage{fancyhdr}
\usepackage{enumerate}
\usepackage{mathtools}
\usepackage{mathdots}
\usepackage{tikz}
\usepackage{graphicx}
\usepackage{makeidx}
\usepackage{color}
\usepackage{caption}
\usepackage{float}
\restylefloat{table}
\usepackage{bbm}
\usepackage{bm}
\usepackage{xcolor}
\usepackage{titlesec}
\usepackage[linesnumbered,ruled]{algorithm2e}
\usepackage{enumitem, color, amssymb}
\usepackage{subcaption}
\usepackage[square,numbers]{natbib}
\setcitestyle{authoryear,open={(},close={)}}

\titleformat*{\section}{\LARGE\fontfamily{ptm}\selectfont}
\titleformat*{\subsection}{\Large\fontfamily{ptm}\selectfont}

\title{A Bayesian framework for case-cohort Cox regression: application to dietary epidemiology}
\author{Andrew Yiu%
  \thanks{Electronic address: \texttt{ahmy2cam@protonmail.com}; Corresponding author}\hspace{.2cm}$^{a}$ }
  \author{Robert J.B. Goudie$^{a}$}
  \author{Stephen J. Sharp$^{b}$}
  \author{Paul J. Newcombe$^{a}$}
  \author{Brian D.M. Tom$^{a}$}
\affil{$^{a}$MRC Biostatistics Unit, University of Cambridge, UK\\
$^{b}$MRC Epidemiology Unit, University of Cambridge, UK}
\date{\today}

\newtheoremstyle{mytheoremstyle} % name
        {\topsep}                    % Space above
        {\topsep}                    % Space below
        {\itshape\fontfamily{ptm}\selectfont}                   % Body font
        {}                           % Indent amount
        {\fontfamily{ptm}\selectfont\scshape}                   % Theorem head font
        {:}                          % Punctuation after theorem head
        {.5em}                       % Space after theorem head
        {}  % Theorem head spec (can be left empty, meaning ‘normal’)
\theoremstyle{mytheoremstyle}

\newtheorem{assumption}{Assumption}

\begin{document}

{\fontfamily{qtm}\selectfont

\maketitle

\newcommand{\iid}{\stackrel{iid}{\sim}}
\newcommand{\fnhat}{\hat{F}_{n}}
\newcommand{\fhhat}{\hat{f}_{h}}
\newcommand{\gnhat}{\hat{G}_{n}}
\newcommand{\thehat}{\hat{\theta}_{n}}
\newcommand{\sumi}{\sum\limits_{i=1}^n}
\newcommand{\sumj}{\sum\limits_{j=1}^N}
\newcommand{\vars}[1]{#1_{1}, \dots ,#1_{n}}
\newcommand{\varsk}[2]{#1_{1}, \dots ,#1_{#2}}
\newcommand{\GG}{\mathcal{G}}
\newcommand{\zi}{z_{i}}
\newcommand\floor[1]{\lfloor#1\rfloor}
\newcommand\ceil[1]{\lceil#1\rceil}
\newcommand\independent{\protect\mathpalette{\protect\independenT}{\perp}}
\def\independenT#1#2{\mathrel{\rlap{$#1#2$}\mkern2mu{#1#2}}}
\newcommand{\gap}{

\vspace{3 mm} \noindent}
\newcommand{\smallg}{

\vspace{1 mm} \noindent}
\newcommand{\real}{\mathbb{R}}
\newcommand{\Proof}{\textit{Proof. }}

\newcommand{\Thm}[1]{\textbf{Theorem #1}}
\newcommand{\Prop}[1]{\textbf{Proposition #1   }}
\newcommand{\Exa}[1]{\textbf{Example #1   }}
\newcommand{\Def}[1]{\textbf{Definition #1   }}
\newcommand{\Lem}[1]{\textbf{Lemma #1   }}
\newcommand{\Cor}[1]{\textbf{Corollary #1   }}
\newcommand{\clg}[1]{\lceil{#1} \rceil}
\newcommand{\intinfx}[1]{\int_{-\infty}^{\infty} #1 \text{ }dx}
\newcommand{\intinf}[2]{\int_{-\infty}^{\infty} #1 \text{ }d#2}
\newcommand{\intx}[3]{\int_{#1}^{#2} #3 dx}
\newcommand{\bo}[1]{\textbf{#1}}
\newcommand{\kh}{k_{h}}
\newcommand{\bigb}[2]{\left(\frac{#1}{#2}\right)}
\newcommand{\E}{\mathbb{E}}
\def\T{{ \mathrm{\scriptscriptstyle T} }}
\newcommand{\begineq}[1]{\begin{equation*}
    \begin{split}
    #1
    \end{split}
\end{equation*}}
\newcommand{\begineqn}[1]{\begin{equation}
    \begin{split}
    #1
    \end{split}
\end{equation}}
\newcommand{\dist}{\xrightarrow{d}}
\newcommand{\Ysbar}{\bar{Y}_{S}}

\begin{abstract}
The case-cohort study design bypasses resource constraints by collecting certain expensive covariates for only a small subset of the full cohort. Weighted Cox regression is the most widely used approach for analysing case-cohort data within the Cox model, but is inefficient. Alternative approaches based on multiple imputation and nonparametric maximum likelihood suffer from incompatibility and computational issues respectively. We introduce a novel Bayesian framework for case-cohort Cox regression that avoids the aforementioned problems. Users can include auxiliary variables to help predict the unmeasured expensive covariates with a prediction model of their choice, while the models for the nuisance parameters are nonparametrically specified and integrated out. Posterior sampling can be carried out using procedures based on the pseudo-marginal MCMC algorithm. The method scales effectively to large, complex datasets, as demonstrated in our application: investigating the associations between saturated fatty acids and type 2 diabetes using the EPIC-Norfolk study. As part of our analysis, we also develop a new approach for handling compositional data in the Cox model, leading to more reliable and interpretable results compared to previous studies. The performance of our method is illustrated with extensive simulations. The code used to produce the results in this paper can be found at \href{https://github.com/andrewyiu/bayes_cc}{https://github.com/andrewyiu/bayes\_cc}.
\end{abstract}

\section{Introduction}
\label{sec:intr}

The case-cohort study design \citep{Prentice86} is an increasingly  common approach for studying prospective epidemiological associations. Time and cost constraints, as well as concerns over the wastage of valuable biological material \citep{Borgan17}, can render it infeasible to obtain certain covariates on a full cohort. The case-cohort design circumvents this issue by restricting complete covariate measurements to a randomly sampled subcohort along with all remaining incident cases, allowing one to efficiently target the quantities of interest while retaining identifiability. An advantage over the similarly motivated nested case-control design \citep{Thomas77} is the ability to reuse the subcohort for multiple endpoints \citep{Kulathinal06}.

Existing proposals for analysing case-cohort data are mostly based on the Cox proportional hazards model \citep{Cox72}, although other models have been considered \citep[e.g.][]{Lu06, Zeng14, Steingrimsson17}. The most widely used approach is weighted Cox regression, motivated by the intuition that the oversampling of cases can be balanced by an appropriate overweighting of the subcohort controls. The methods of \citet{Prentice86} and \citet{Barlow94} are the most commonly applied \citep{Sharp14}. In both proposals, cases sampled outside of the subcohort enter into the analysis only at their respective failure times, allowing for the partially collected covariates---referred to as \textit{expensive} covariates hereafter---to be time-dependent.
%the motivation behind the case-cohort design suggests that repeated measurements of the expensive covariates across time are likely to be infeasible; in which case, the assumption of time-independence is required regardless.

Assuming time-independence permits more efficient weighting approaches. \citet{Kalbfleisch88} and \citet{Chen99} proposed weighting schemes based on inverse probability weighting and post-stratification respectively. However, neither approach can make use of potentially available information on the unsampled controls, such as auxiliary variables and censoring times. \citet{Borgan00} suggested several methods to address this issue, one of which was later augmented by \citet{Kulich04} to increase efficiency. Yet, weighted Cox estimators cannot be fully efficient; \cite{Nan04a} studied the semiparametric efficiency bound for the problem and quantified the amount of efficiency lost. It is unclear whether estimators that achieve the bound can be constructed in general. %\citet{Qi05} and \citet{Luo09} studied augmented weighted estimators when covariates are missing at random.

Alternatives to weighted Cox regression have been proposed that use the full cohort data more efficiently and avoid the potential instability of inverse probability weights. \citet{Keogh13} described how multiple imputation can be applied to the problem, treating the expensive covariates for unsampled individuals as missing data. This requires a conditional imputation model for the expensive covariates given all observed variables, including the event time and case indicator. Care is required to avoid incompatibility issues with the proportional hazards model: \citet{Keogh13} implemented the imputation with either a simplistic generalized linear model, or with rejection sampling using a preliminary marginal model. Full likelihood methods have assumed that the censoring mechanism is ignorable given the observed data. Nonparametric maximum likelihood estimation with the EM-algorithm was proposed by \citet{Scheike04}, later extended by \citet{Zeng14} to include auxiliary variables and shown to be semiparametric efficient. However, computation is numerically unstable for more than three continuous auxiliary variables. \citet{Kulathinal06} considered Bayesian analysis with data augmentation \citep{Tanner87}, specifying a fully parametric form for the baseline cumulative hazard function. 

We introduce a novel Bayesian framework for case-cohort Cox regression under the ignorable censoring assumption stated earlier; time-independence will be also be assumed since it is sufficient for our application and simplifies the descriptions, but we will discuss how this can be relaxed. The basic procedure is carried out in two stages. First, we obtain the posterior of the conditional distribution of the expensive covariates given the fully observed covariates using only the data from individuals with complete measurements---we refer to this as the restricted posterior. Samples from this restricted posterior serve as inputs to a pseudo-marginal Metropolis-Hastings algorithm \citep{Andrieu09}. This procedure yields the interpretation of using a likelihood function equal to the average of a set of Cox partial likelihoods, each computed from a dataset formed from the original with a different instance of imputed values for the missing expensive covariates. In this regard, our method shares a conceptual similarity with multiple imputation, but is fully Bayesian and is automatically free of incompatibility issues with the Cox model. For large and moderate-dimensional datasets, we also propose extensions to the method based on modified versions of the correlated pseudo-marginal algorithm \citep{Deligiannidis18} that facilitate faster mixing. 

Unlike \citet{Kulathinal06}, who require a fully specified joint model for the expensive and fully observed covariates, we allow for the (nuisance) marginal distribution of the fully observed covariates to be ignored. Moreover, our model for the baseline cumulative hazard is nonparametrically specified and integrated out; this obviates sampling a potentially high-dimensional (or even infinite-dimensional) parameter, and leads to more robust inference for the log-hazard ratio than using a parametric model specification. With no auxiliary variables, and a discrete model for the expensive covariates, the likelihood reduces to the nonparametric likelihood used by \citet{Scheike04}. When auxiliary variables are available, the conditional model for the expensive covariates can be arbitrarily specified, without the three dimensional covariate ceiling of the \citet{Zeng14} kernel estimation approach.

In Section \ref{sec:meth}, we introduce our method in a general setting, and propose modifications to the basic algorithm that facilitate improved mixing. Simulations comparing the performance of our approach to previous proposals are presented in Section \ref{sec::sim}. In Section \ref{app}, we apply our method to the EPIC-Norfolk study with the objective of investigating the associations between individual plasma phospholipid saturated fatty acids and incident type 2 diabetes. A challenging aspect is incorporating the compositional fatty acid data into the Cox model. Previous studies treated the proportions as absolute measurements, and used them directly. On the other hand, we first apply the additive logratio transformation \citep{Aitchison82} to the data. We discuss how this produces more reliable and interpretable results.

\section{Bayesian case-cohort Cox regression}
\label{sec:meth}

\subsection{Notation and background}
\label{sec:setup}

First, consider the Cox proportional hazards model \citep{Cox72} for complete data. Let $D^{0}=(Y, \Delta, Z, W)$, where $Y=\min(T, C)$, $T$ and $C$ denote the failure time and right-censoring time respectively, $\Delta = I(T \leq C)$ and $(Z,W) \in \mathbb{R}^{m}$ is a vector of time-independent covariates---later on, there is a probability that $Z$ is unobserved. The conditional hazard function of $T$ given $(Z,W)$ is $\lambda(t) = \exp(\beta_{1}^{\T}Z+\beta_{2}^{\T}W)\lambda_{0}(t)$, where $\beta = (\beta_{1}, \beta_{2})$ is the log-hazard ratio describing the effects of the covariates, and $\lambda_{0}(t)$ is the baseline hazard function. Let $\Lambda_{0}(t) = \int_{s=0}^t \lambda_{0}(s)\, ds$ be the baseline cumulative hazard function. Suppose we observe an independent and identically distributed sample $D^{0}_{i} = (Y_{i}, \Delta_{i}, Z_{i}, W_{i})$ $(i=1, \ldots, n)$ and let $R_{i}(t) = I(t \leq Y_{i})$ be the at-risk indicator at time $t$ for individual $i$. Assuming that $T$ and $C$ are conditionally independent given $(Z,W)$, the parameter $\beta$ can be estimated by maximizing the Cox partial likelihood \citep{Cox72}
\[
    \prod_{i = 1}^n \left\{\frac{\exp{(\beta_{1}^{\T}Z_{i}}+\beta_{2}^{\T}W_{i})}{\sum_{j=1}^n R_{j}(T_{i})\exp{(\beta_{1}^{\T}Z_{j}+\beta_{2}^{\T}W_{j})} }\right\}^{\Delta_{i}}.
\]
In the presence of ties, the above expression takes the Breslow form of the partial likelihood \citep{Breslow72}, which is the form we will use for the whole of this paper.

Suppose now that the covariates $Z_{i}$---which we will refer to as the expensive covariates---are measured for only a random subset of the cohort. Suppose also that we observe an independent and identically distributed sample $X_{i}$ $(i=1, \ldots, n)$ of auxiliary covariates that can be used to predict the unmeasured values of $Z$. More explicitly, we observe $D_{i} = (Y_{i}, \Delta_{i}, A_{i}Z_{i}, A_{i}, W_{i}, X_{i})$ $(i=1, \ldots, n)$, where $A_{i}$ is a binary variable indicating whether the expensive covariates for individual $i$ have been measured, and the other variables are defined as before. In a standard case-cohort design, $A_{i} = 1$ if individual $i$ is a case, or a control sampled into the subcohort. Let $\mathcal{S} = \{i: A_{i} = 1\} \subset \{1, \ldots, n\}$ denote the set of individuals with measured $Z_{i}$, and let $\bar{\mathcal{S}} = \{1, \ldots, n\} \setminus \mathcal{S}$. We will make use of the shorthand notation of indexing by sets, e.g. $X_{\mathcal{S}} = \{X_{i}: i \in \mathcal{S}\}$. 

We make the following assumptions:
\begin{assumption}
\label{assumption1}
For each $i =1,\ldots, n$, $C_{i}$ is independent of $(T_{i},Z_{i})$ given $(W_{i},X_{i})$.
\end{assumption}
\begin{assumption}
\label{assumption2}
The vector $(A_{1}, \ldots, A_{n})$ is independent of $(Z_{1}, \ldots, Z_{n})$ given $\{(Y_{j},\Delta_{j},W_{j},X_{j}): j=1,\ldots, n\}$.
\end{assumption}
Assumption~\ref{assumption1} strengthens the conditional independence assumption for full-data Cox regression, requiring further that $C_{i}$ be independent of $Z_{i}$ given $(W_{i},X_{i})$ for each $i=1, \ldots, n$. This will hold, for example, if the censoring is administrative. Assumption~\ref{assumption2} is guaranteed to hold for standard case-cohort studies since the subcohort selection mechanism is known by design, and is either fully randomized or stratified on the baseline covariates $X_{i}$. %CHANGE LAST SENTENCE

\subsection{Model and inference}
\label{sec:proposal}
%We begin by motivating our method and model specification, providing a formal justification with Theorem 1 later on. 
Under the general set-up described in  \S\ref{sec:setup}, the likelihood function for the data $D_{1}, \ldots, D_{n}$ is equal to
\begin{equation} \label{eqn:like}
\begin{split}
    \left[ \prod_{i\in \mathcal{S}} \{\exp(\beta_{1}^{\T}Z_{i}+\beta_{2}^{\T}W_{i})\lambda_{0}(Y_{i})\}^{\Delta_{i}}\exp\left\{-e^{\beta_{1}^{\T}Z_{i}+\beta_{2}^{\T}W_{i}}\Lambda_{0}(Y_{i})\right\}p(Z_{i} \mid W_{i},X_{i})\right] \\ \left[\prod_{j\in \bar{\mathcal{S}}}\int\exp\left\{-e^{\beta_{1}^{\T}z_{j}+\beta_{2}^{\T}W_{j}}\Lambda_{0}(Y_{j})\right\}p(z_{j} \mid W_{j},X_{j}) dz_{j} \right]
    \end{split}
\end{equation}
multiplied by
\begin{equation} \label{eqn:nuis}
\begin{split}
    \left\{\prod_{k=1}^{n} p(C_{k} \mid W_{k}, X_{k})^{1-\Delta_{k}}\mathbb{P}(C_{k} \geq Y_{k} \mid W_{k}, X_{k})^{\Delta_{k}}p(W_{k}, X_{k})\right\}\\
    p(A_{1}, \ldots, A_{n} \mid \{(Y_{j},\Delta_{j},W_{j},X_{j}): j=1,\ldots, n\}).
    \end{split}
\end{equation}
This is derived by taking the full likelihood for the Cox model with complete data \citep[p.425]{vanderVaart98} and integrating out the missing expensive covariates $\{Z_{j}:\,j \in \bar{\mathcal{S}}\}$. In this section, we will describe our model restrictions for the different terms in the likelihood, and explain how to carry out inference on the hazard ratio.

The baseline cumulative hazard function $\Lambda_{0}$ is set to be a step function with jumps only at the failure times. Let $\Delta\Lambda_{0}(Y_{i})$ denote the jump size of $\Lambda_{0}$ at $Y_{i}$ for $\Delta_{i}=1$. Then, the baseline hazard $\lambda_{0}(Y_{i})$ equals $\Delta\Lambda_{0}(Y_{i})$ if $\Delta_{i} = 1$, and $0$ otherwise, and $\Lambda_{0}(t) = \sum_{i: \Delta_{i} = 1, Y_{i} \leq t} \Delta\Lambda_{0}(Y_{i})$. This idea was introduced by \citet{Breslow72} to motivate both the Cox partial likelihood estimator from a nonparametric maximum likelihood perspective, and the Breslow estimator of the baseline cumulative hazard function. \citet{Scheike04} and \citet{Zeng14} extended this approach for case-cohort data. 

We specify a Bayesian bootstrap prior for $\Lambda_{0}$
\begin{equation*}
    p(\Lambda_{0}) \propto \prod_{i: \Delta_{i} = 1} \Delta\Lambda_{0}(Y_{i})^{-1}.
\end{equation*}
For complete data, \citet{Kim03} referred to this as the ``Poisson form Bayesian bootstrap'' and showed that the resulting inference for $\beta$ is equivalent to Bayesian analysis with the Cox partial likelihood. We will see that a similar phenomenon arises with case-cohort data. \citet{Kalbfleisch78} and \citet{Sinha03} motivated this prior by considering the limit of a sequence of gamma process priors that become progressively more noninformative. This is similar to how the original Bayesian bootstrap \citep{Rubin81} can be motivated by considering the noninformative limit of a sequence of Dirichlet process priors.  %add explanation and justifications for using the Bayesian bootstrap prior, including reference to the Kim and Lee paper and reference to Asymptotic Statistics.

For the terms of the form $p(Z \mid W, X)$ in (\ref{eqn:like}), we require a regression model for the expensive covariates $Z$ given the fully observed covariates $(W,X)$. This will be used to predict the missing expensive covariate values and its specification is left to the user. We denote the parameter of this model by $\gamma$, which can be infinite-dimensional. The priors for $\beta$ and $\gamma$ are also left to the user, aside from the requirement of joint prior independence of $\Lambda_{0}$, $\beta$, and $\gamma$.

In (\ref{eqn:nuis}), we set the models for the censoring $p(C_{k} \mid W_{k}, X_{k})$, the fully observed covariates $p(W_{k}, X_{k})$ and the selection $p(A_{1}, \ldots, A_{n} \mid \{(Y_{j},\Delta_{j},W_{j},X_{j}): j=1,\ldots, n\})$ to be a priori independent of $(\Lambda_{0}, \beta, \gamma)$. Thus, (\ref{eqn:nuis}) will drop out of the subsequent analysis and no further specification of these models is needed.

It follows that the posterior for $( \Lambda_{0},\beta, \gamma)$ given $D_{1}, \ldots, D_{n}$ is proportional to
\begin{equation}
\begin{split}
\label{eqn:post}
    \left[ \prod_{i\in \mathcal{S}} \exp(\beta_{1}^{\T}Z_{i}+\beta_{2}^{\T}W_{i})^{\Delta_{i}}\exp\left\{-e^{\beta_{1}^{\T}Z_{i}+\beta_{2}^{\T}W_{i}}\Lambda_{0}(Y_{i})\right\}p(Z_{i} \mid W_{i},X_{i}, \gamma)\right]\\ \left[\prod_{j\in \bar{\mathcal{S}}}\int\exp\left\{-e^{\beta_{1}^{\T}z_{j}+\beta_{2}^{\T}W_{j}}\Lambda_{0}(Y_{j})\right\}p(z_{j} \mid W_{j},X_{j}, \gamma) dz_{j} \right] p(\gamma)p(\beta).
    \end{split}
\end{equation}
Let \begin{equation} \label{eqn:partpost}
    p(\gamma \mid D_{\mathcal{S}})  \propto \left[ \prod_{i\in \mathcal{S}} p(Z_{i} \mid W_{i},X_{i}, \gamma)\right] p(\gamma)
\end{equation}
be the posterior for $\gamma$ given only the data for individuals in $\mathcal{S}$---the set of individuals with measured $Z_{i}$. We refer to this as the \textit{restricted posterior} of $\gamma$. By integrating (\ref{eqn:post}) with respect to $\Lambda_{0}$, applying Fubini's theorem to exchange the order of integration with the missing covariates, and then integrating with respect to $\gamma$, we find that 
\begin{equation}
\label{eqn:margpost}
\begin{split}
    p(\beta \mid D_{1}, \ldots, D_{n}) \propto \int \prod_{k = 1}^n \left\{\frac{\exp{(\beta_{1}^{\T}z_{k}}+\beta_{2}^{\T}W_{k})}{\sum_{l=1}^n R_{l}(T_{k})\exp{(\beta_{1}^{\T}z_{l}+\beta_{2}^{\T}W_{l})} }\right\}^{\Delta_{k}} \left[\prod_{i\in \mathcal{S}} \delta\{z_{i} = Z_{i}\}dz_{i}\right] \\ \left[\prod_{j\in \bar{\mathcal{S}}} p(z_{j} \mid W_{j},X_{j}, \gamma) dz_{j} \right] p(\gamma \mid D_{\mathcal{S}})  d\gamma \,p(\beta)
    \end{split}
\end{equation}
where $\delta\{\cdot\}$ is the Dirac delta function (a more detailed derivation can be found in \autoref{appen::deriv}). Thus, the posterior of $\beta$ is proportional to the prior of $\beta$ multiplied by the Cox partial likelihood averaged across the restricted posterior predictive distribution of the missing covariates. 

Although this averaged Cox partial likelihood is probably intractable, it is generally possible to draw values of the missing covariates from the restricted posterior predictive distribution, either exactly or by MCMC methods. This provides us with a computational strategy to sample from the marginal posterior of $\beta$ using a pseudo-marginal Metropolis-Hastings algorithm \citep{Andrieu09}. Let $B$ be a positive integer (the choice of which is suggested below). Define the distribution of a $B \times |\bar{\mathcal{S}}|$ random variable $Z^{\text{mis}}$ by
\begin{equation} \label{eqn::zmis}
    p(z^{\text{mis}} \mid W_{\bar{\mathcal{S}}},X_{\bar{\mathcal{S}}}, D_{\mathcal{S}}) =  \prod_{b=1}^{B} \int  \prod_{j \in \bar{\mathcal{S}}} p(z_{j}^{(b)} \mid W_{j}, X_{j},\gamma_{b})p(\gamma_{b} \mid D_{\mathcal{S}})d\gamma_{b},
\end{equation}
where $\{z_{j}^{(b)}:\,  j \in \bar{\mathcal{S}},\, b = 1, \ldots, B\}$ are the components of $z^{\text{mis}}$. We can sample $Z^{\text{mis}}$ as follows: draw $B$ independent values $\gamma_{1}, \ldots, \gamma_{B}$ from the restricted posterior (\ref{eqn:partpost}), and for each $b=1, \ldots, B$ and each $j \in \bar{\mathcal{S}}$, draw from $p(z \mid W_{j},X_{j}, \gamma_{b})$; $Z^{\text{mis}}$ takes the value of the set of imputed covariates. By combining $Z^{\text{mis}}$ with the measured values of $Z$, this procedure yields $B$ datasets with complete covariate measurements. Define the function $h$ by the mean of the partial likelihood functions across all datasets:
\begin{equation*}
    h(\beta, Z^{\text{mis}}) = B^{-1}\sum_{b=1}^{B} \left[ \prod_{k: \Delta_{k} = 1} \frac{\exp(\beta_{1}^{\T}Z_{k}+\beta_{2}W_{k})}{\sum_{l=1}^{n}R_{l}(Y_{k})\exp(\beta_{1}^{\T}Z_{l}^{(b)}+\beta_{2}W_{l})}\right]
\end{equation*}
where $Z_{l}^{(b)}$ is the expensive covariate for individual $l$ in the $b$-th imputed dataset.

\begin{algorithm}[!h]
\caption{Sampling from the marginal posterior of $\beta$} \label{al1}
\begin{tabbing}
    \enspace Input initial parameter value $\beta^{(0)}$.\\
   \enspace Draw $Z^{\text{mis}}_{(0)}$ from $p(z^{\text{mis}} \mid W_{\bar{\mathcal{S}}}, X_{\bar{\mathcal{S}}}, D_{\mathcal{S}})$ (\ref{eqn::zmis}).\\
   \enspace For $r=1$ to $r = N$ \\
   \qquad (a) Propose $\Tilde{\beta}$ from $q(\beta \mid \beta^{(r-1)})$.\\
   \qquad (b) Draw $\Tilde{Z}^{\text{mis}}$ from $p(z^{\text{mis}} \mid W_{\bar{\mathcal{S}}}, X_{\bar{\mathcal{S}}}, D_{\mathcal{S}})$.\\
   \qquad (c) With probability  $\min\left\{1, \frac{q( \beta^{(r-1)} \mid \Tilde{\beta}) p(\Tilde{\beta})h(\Tilde{\beta}, \Tilde{Z}^{\text{mis}})}{q(\Tilde{\beta} \mid \beta^{(r-1)})p(\beta^{(r-1)}) h(\beta^{(r-1)}, Z_{(r-1)}^{\text{mis}})} \right\}$, set $ \beta^{(r)} =\Tilde{\beta}$ and  $Z_{(r)}^{\text{mis}} = \Tilde{Z}^{\text{mis}}$. \\
    \qquad Otherwise, set $ \beta^{(r)} =\beta^{(r-1)}$ and  $Z^{\text{mis}}_{(r)} = Z^{\text{mis}}_{(r-1)}$. \\
    \enspace Output $(\beta^{(1)}, \ldots, \beta^{(N)})$.
\end{tabbing}
\vspace*{-6pt}
\end{algorithm}

Let $q(\Tilde{\beta} \mid \beta)$ be a user-specified proposal distribution for $\beta$. Algorithm~\ref{al1} describes the basic template for sampling from the marginal posterior of $\beta$. The algorithm can be viewed as a Metropolis-Hastings algorithm for the augmented parameter $(\beta, z^{\text{mis}})$ with proposal distribution $q^{*}(\Tilde{\beta}, \Tilde{Z}^{\text{mis}} \mid \beta, z^{\text{mis}}) = q(\Tilde{\beta} \mid \beta) p(\Tilde{z}^{\text{mis}} \mid W_{\bar{\mathcal{S}}}, X_{\bar{\mathcal{S}}}, D_{\mathcal{S}})$. The acceptance probability for the $r$-th iteration with proposal $(\Tilde{\beta}, \Tilde{Z}^{\text{mis}})$ and current value $(\beta^{(r-1)}, Z_{(r-1)}^{\text{mis}})$ can now be written as
$$\min\left\{1, \frac{q^{*}(\beta^{(r-1)}, Z^{\text{mis}}_{(r-1)} \mid \Tilde{\beta}, \Tilde{Z}^{\text{mis}})  p(\Tilde{\beta})p(\Tilde{Z}^{\text{mis}} \mid W_{\bar{\mathcal{S}}},X_{\bar{\mathcal{S}}}, D_{\mathcal{S}})h(\Tilde{\beta}, \Tilde{Z}^{\text{mis}})}{q^{*}(\Tilde{\beta}, \Tilde{Z}^{\text{mis}} \mid \beta^{(r-1)}, Z^{\text{mis}}_{(r-1)}) p(\beta^{(r-1)})p(Z^{\text{mis}}_{(r-1)} \mid W_{\bar{\mathcal{S}}}, X_{\bar{\mathcal{S}}}, D_{\mathcal{S}}) h(\beta^{(r-1)}, Z_{(r-1)}^{\text{mis}})} \right\}.$$ Thus, Algorithm~\ref{al1} converges to stationarity with an invariant distribution function proportional to $p(\beta)p(z^{\text{mis}} \mid W_{\bar{\mathcal{S}}}, X_{\bar{\mathcal{S}}},D_{\mathcal{S}})h(\beta, z^{\text{mis}})$. By construction, the expectation of $h(\beta, Z^{\text{mis}})$ with respect to $p(z^{\text{mis}} \mid W_{\bar{\mathcal{S}}}, X_{\bar{\mathcal{S}}}, D_{\mathcal{S}})$ is proportional to $p(D_{1}, \ldots, D_{n} \mid \beta)$ in $\beta$; the marginal invariant distribution of $\beta$ is therefore equal to the true marginal posterior. If MCMC is required to draw restricted posterior values of $\gamma$, it is straightforward to modify Algorithm~\ref{al1} to sample the further augmented parameter $(\beta, z^{\text{mis}}, \gamma_{1}, \ldots, \gamma_{B})$. % by adding $\prod_{b=1}^{B} p(\gamma_{b} \mid D_{\mathcal{S}})$ terms to the acceptance ratio.

Since Algorithm~\ref{al1} is a pseudo-marginal algorithm that uses an average of unbiased estimators (as opposed to a particle filter), and computation time scales roughly linearly in $B$, the results of \citet{Sherlock17} suggest that the optimal computational tradeoff between number of iterations $N$ and number of estimators $B$ is achieved by setting $B=1$. If parallel computing is available with negligible overheads, $B$ should be set equal to the number of available cores, so that the $B$ partial likelihood functions are computed in parallel.

\subsection{Modifications to improve mixing} \label{sec:corr}

For large datasets with moderate to high dimensional covariates, such as our application in \S\ref{app}, Algorithm~\ref{al1} may not be sufficient to ensure good mixing. In this section, we describe how improved mixing can be attained.

The correlated pseudo-marginal algorithm \citep{Deligiannidis18} improves on the efficiency of the standard pseudo-marginal algorithm by correlating the current and proposed values of the variables that are used to obtain the estimate of the likelihood factor ($Z^{\text{mis}}$ in our set-up). However, this method requires the distribution of these variables to be inverted into a standard  multivariate normal distribution; for the restricted posterior predictive distribution of $Z^{\text{mis}}$ given by (\ref{eqn::zmis}), this will generally be impossible in practice due to intractability.

We solve this by instead considering the restricted posterior predictive distribution of $Z^{\text{mis}}$ \textit{conditional} on $\gamma_{1}, \ldots, \gamma_{B}$. In equation (\ref{eqn::zmis}), the factors of the form $p(z_{j}^{(b)} \mid W_{j}, X_{j},\gamma^{(b)})$ are user-specified probability density/mass functions. Generally, this means that we can analytically or numerically evaluate a deterministic function $\varphi$ such that $\varphi(U,W_{\bar{\mathcal{S}}}, X_{\bar{\mathcal{S}}}, \gamma_{1}, \ldots, \gamma_{B})$ has the distribution of $Z^{\text{mis}}$, where $U \sim \mathcal{N}(0_{M}, I_{M})$ for $M= B \times \bar{\mathcal{S}}$, independent of $\gamma_{1}, \ldots, \gamma_{B}$. This motivates Algorithm~\ref{al2}, a modified version of the correlated pseudo-marginal algorithm in which the set of parameters is augmented by $\gamma_{1}, \ldots, \gamma_{B}$, and the values of $U$ are correlated to the level determined by $\rho \in (-1,1)$. When $\rho =0$, Algorithm~\ref{al2} is equivalent to Algorithm~\ref{al1}. Increasing $\rho$ leads to higher acceptance probabilities but slower exploration of the parameter space; the value can be tuned accordingly. We justify the algorithm in \autoref{appen:alg2}.

\begin{algorithm}[!h]
\caption{Correlated sampling algorithm} \label{al2}
\begin{tabbing}
    \enspace Input initial parameter value $\beta^{(0)}$\\
   \enspace Draw $U^{(0)} \sim \mathcal{N}(0_{M}, I_{M})$.\\
   \enspace Draw i.i.d. $\gamma^{(0)}_{1}, \ldots, \gamma^{(0)}_{B} \sim p(\gamma \mid D_{\mathcal{S}})$. \\
   \enspace Compute $Z^{\text{mis}}_{(0)}=\varphi(U^{(0)}, W_{\bar{\mathcal{S}}}, X_{\bar{\mathcal{S}}}, \gamma^{(0)}_{1}, \ldots, \gamma^{(0)}_{B})$.\\
   \enspace For $r=1$ to $r = N$ \\
   \qquad (a) Draw a proposal $\Tilde{\beta}$ from $q(\beta \mid \beta^{(r-1)})$.\\
   \qquad (b) Draw i.i.d. $\Tilde{\gamma}_{1}, \ldots, \Tilde{\gamma}_{B} \sim p(\gamma \mid D_{\mathcal{S}})$.\\
   \qquad (c) Draw $\varepsilon \sim  \mathcal{N}(0_{M}, I_{M})$ and set $\Tilde{U} = \rho U^{(r-1)} + \sqrt{(1-\rho^{2})} \varepsilon$. \\
   \qquad (d) Compute $\Tilde{Z}^{\text{mis}} = \varphi(\Tilde{U}, W_{\bar{\mathcal{S}}}, X_{\bar{\mathcal{S}}},\Tilde{\gamma}_{1}, \ldots, \Tilde{\gamma}_{B})$.\\
   \qquad (e) With probability  $\min\left\{1, \frac{q( \beta^{(r-1)} \mid \Tilde{\beta}) p(\Tilde{\beta})h(\Tilde{\beta}, \Tilde{Z}^{\text{mis}})}{q(\Tilde{\beta} \mid \beta^{(r-1)})p(\beta^{(r-1)}) h(\beta^{(r-1)}, Z_{(r-1)}^{\text{mis}})} \right\}$, set $(\beta^{(r)}, U^{(r)}) = (\Tilde{\beta}, \Tilde{U})$. \\
    \qquad Otherwise, set $(\beta^{(r)}, U^{(r)}) =(\beta^{(r-1)}, U^{(r-1)})$\\
    \enspace Output $(\beta^{(1)}, \ldots, \beta^{(N)})$.
\end{tabbing}
\vspace*{-6pt}
\end{algorithm}

If this is insufficient to ensure adequate mixing, we can correlate $\gamma_{1}, \ldots, \gamma_{B}$ as well. In the case where the restricted posterior $p(\gamma \mid D_{\mathcal{S}})$ admits an analytic expression, it is straightforward to extend Algorithm~\ref{al2} by replacing step (b) with a correlated proposal using the normal inversion strategy employed for $Z^{\text{mis}}$. We take this approach in \S\ref{app}, albeit only for a subparameter of $\gamma$. Otherwise, we can sample $\gamma_{1}, \ldots, \gamma_{B}$ using a Metropolis-Hastings algorithm with a proposal distribution chosen to induce a suitable level of correlation.

\section{Simulation study} \label{sec::sim}

In this section, we assess our proposal by comparing its performance with existing methods by \citet{Prentice86}, \citet{Kalbfleisch88} and \citet{Chen99}. Since these methods are unable to incorporate auxiliary covariates to improve the prediction of the missing expensive covariates, we considered the special case where there are no auxiliary covariates to enable direct comparisons.

Failure times were independently and identically generated for a full cohort size of $n=2000$ using a Weibull baseline hazard function
\begin{equation*}
    \lambda(t) = \exp(\beta_{0}Z)\eta \nu t^{\nu-1},
\end{equation*}
where $\beta_{0}$ is the target parameter. The expensive covariate $Z$ was generated from $\mathcal{N}(0,1)$. The censoring times took the value $3$ with probability $0.2$, and were otherwise uniformly distributed between $0$ and $3$. The sets of values of $(\beta_{0}, \eta, \nu)$, with $\beta \in \{-0.3,0,0.3\}$, were chosen such that the average proportion of cases (approximately 4\%) roughly corresponded to that of the application. The subcohort sampling proportion $p = 0.04$ was chosen similarly.

\iffalse
The approach by \citet{Scheike04} estimates $\beta_{0}$ by nonparametric maximum likelihood. Similar to our proposal, the baseline cumulative hazard function is modeled as a step function which jumps only at the failure times. Further discretization is achieved by assuming the distribution of $Z$ is supported only on the observed values. We compute the estimator with the EM Algorithm \citep{Dempster77}; one alternates between taking the expectation of the complete data log-likelihood given the current values of the parameters, and maximizing this function to update the parameter values. The initial values for $\beta$ and $\Lambda_{0}$ were set using a \citet{Chen99} weighted Cox analysis, while the initial distribution for $Z$ was set to be uniform. The algorithm was iterated until consecutive values of $\beta$ differed by less than 0.0001.
\fi

The proposals by \citet{Prentice86}, \citet{Kalbfleisch88} and \citet{Chen99} solve weighted versions of the Cox partial score equation and can be implemented using the R package \texttt{survival}. For our Bayesian method, computation was carried out using Algorithm~\ref{al1}. We specified a Bayesian bootstrap model \citep{Rubin81} for the distribution of $Z$. A new value of $Z^{\text{mis}}$ is proposed as follows: sample a set of probability weights from $\text{Dirichlet}(1, \ldots, 1)$, each corresponding to an observed value of $Z$ in $\mathcal{S}$; conditional on the weights, independently draw each missing covariate from the observed set of $Z$ values. For $\beta$, we specified an improper uniform prior on $\mathbb{R}$ and used a normal random walk proposal: $q(\tilde{\beta} \mid \beta^{(r-1)}) = \mathcal{N}(\beta^{(r-1)}, \sigma^{2})$. For the proposal variance, we used four times the estimated variance of the \citet{Chen99} estimator (using a weighted Cox analysis). With parallel computing, the communication overhead dominated the computation time of the likelihood estimator; thus, the number of estimators $B$ was set to 1. The first 1000 Metropolis-Hastings iterations were discarded, and the subsequent 20000 iterations were used for analysis. We chose the posterior mean as the Bayes point estimator.

Table~\ref{tab:tabtwo} summarizes the performance of the four methods across 2000 Monte Carlo trials. The relative efficiencies were computed by taking the ratio of the mean squared errors relative to the complete data analysis, where information on all variables is available for the full cohort. The coverage properties of the Bayesian method were assessed by examining the proportion of trials where $\beta_{0}$ was contained in the central 95\% posterior credible region. For the remaining procedures, we have reported the coverage from 95\% Wald intervals with robust variance estimates. 

Our proposal substantially outperformed the three weighted Cox approaches in all settings: the Bayes estimator was approximately unbiased with smaller standard deviations, leading to a significant reduction in efficiency loss relative to the complete data analysis. The central posterior credible regions also exhibited frequentist coverage close to nominal levels, improving on the Prentice method in particular. We draw attention to the fact that we have specified a noninformative prior for $\beta$ and a nonparametric model for $Z$ that makes virtually no modeling assumptions. Thus, there is ample scope to make further performance gains if prior substantive knowledge is available. 

We mention also that we implemented the nonparametric maximum likelihood estimator \citep{Scheike04, Zeng14}, which is computed using an EM-algorithm. However, we were unable to obtain numerical convergence for any of the sets of parameter values, so we excluded this estimator from the comparisons.

\begin{table}
\caption{Comparison of log-hazard ratio estimates for 2000 replicates. CL, \citet{Chen99}; KL, \citet{Kalbfleisch88}; ESD, empirical standard deviation; RMSE, root mean squared error; RE, relative efficiency; Cov, coverage. \label{tab:tabtwo}}
\begin{center}
\resizebox{6in}{!}{\begin{tabular}{l |ccccc| ccccc}
& \multicolumn{5}{c|}{$\beta_{0} = -0.3,\, \eta = 0.01, \,\nu = 2.0$} & \multicolumn{5}{c}{$\beta_{0} = -0.3,\, \eta = 0.02, \,\nu = 1.2$}\\ \hline
Estimator & Bias & ESD & RMSE & RE & Cov (\%)& Bias & ESD & RMSE & RE & Cov (\%)\\ \hline
Full & 0.000 & 0.109 & 0.109 & 1.000 & 95.00 & 0.000 & 0.107 & 0.107 & 1.000 & 95.70\\
Bayes & 0.013 & 0.145 & 0.146 & 0.561 & 94.80 & 0.012 & 0.141 & 0.142 & 0.563 & 95.85\\
CL & -0.021 &  0.206 & 0.207 &  0.278 & 94.25 & -0.017 & 0.190 & 0.191 & 0.312 & 94.50\\
KL & -0.021 &  0.206 & 0.207 &  0.278 & 94.25 & -0.017 & 0.190 & 0.191 & 0.312 & 94.50\\
Prentice & -0.012 & 0.202 & 0.202 & 0.293  & 90.15 & -0.010 & 0.186 & 0.187 & 0.325 & 91.20\\[2ex]
\end{tabular}}
\end{center}
\begin{center}
\resizebox{6in}{!}{\begin{tabular}{l |ccccc | ccccc}
& \multicolumn{5}{c|}{$\beta_{0} = 0,\, \eta = 0.01, \,\nu = 2.0$}& \multicolumn{5}{c}{$\beta_{0} = 0,\, \eta = 0.02, \,\nu = 1.2$}\\ \hline
Estimator & Bias & ESD & RMSE & RE & Cov (\%)& Bias & ESD & RMSE & RE & Cov (\%)\\ \hline
Full & 0.002 & 0.115 & 0.115 & 1.000 & 94.40 & 0.002 & 0.113 & 0.113 & 1.000& 95.00\\
Bayes & 0.004 & 0.161 & 0.161 & 0.508 & 94.80 & 0.005 & 0.163 & 0.163 & 0.485 & 95.10\\
CL & 0.003 &  0.194 & 0.194 &  0.352 & 95.25 & 0.003 & 0.181 & 0.181 & 0.394 & 95.75\\
KL & 0.003 &  0.194 & 0.194 &  0.352 & 95.25 & 0.003 & 0.181 & 0.181 & 0.394 & 95.75\\
Prentice & 0.002 & 0.191 & 0.191 & 0.363 & 90.20 & 0.003 & 0.178 & 0.178 & 0.405 & 91.20\\[2ex]
\end{tabular}}
\end{center}
\begin{center}
\resizebox{6in}{!}{\begin{tabular}{l |ccccc | ccccc}
& \multicolumn{5}{c|}{$\beta_{0} = 0.3,\, \eta = 0.01, \,\nu = 2.0$}& \multicolumn{5}{c}{$\beta_{0} = 0.3,\, \eta = 0.02, \,\nu = 1.2$}\\ \hline
Estimator & Bias & ESD & RMSE & RE & Cov (\%)& Bias & ESD & RMSE & RE & Cov (\%)\\ \hline
Full & 0.001 & 0.114 & 0.114 & 1.000 & 94.10 & 0.001 & 0.112 & 0.112 & 1.000& 93.65\\
Bayes & -0.008 & 0.151 & 0.151 & 0.571 & 94.80 & -0.008 & 0.149 & 0.150 & 0.564 & 94.90\\
CL & 0.023 &  0.204 & 0.206 &  0.307 & 93.70 & 0.019 & 0.192 & 0.193 & 0.340 & 94.30\\
KL & 0.023 & 0.204 & 0.206 & 0.307 & 93.70 & 0.019 & 0.192 & 0.193 & 0.340 & 94.30\\
Prentice & 0.014 & 0.201 & 0.202 & 0.319 & 89.45 & 0.012 & 0.188 & 0.189 & 0.355 & 90.65\\[2ex]
\end{tabular}}
\end{center}
\end{table}

\section{Application to the EPIC-Norfolk study} \label{app}

\subsection{Study design and data preparation}\label{studes}

We apply our methodology to investigate the associations between individual saturated fatty acids and incident type 2 diabetes, using data from the European Prospective Investigation into Cancer and Nutrition (EPIC)-Norfolk study. The original cohort study included 25,639 men and women aged 40 to 79. Between 1993 and 1997, all participants were invited to undergo a baseline health check, during which anthropometric measurements and blood samples were taken by trained nurses. Participants were also required to complete a health and lifestyle questionnaire. Follow-up concluded on 31\textsuperscript{st} December 2007; the follow-up time for each participant was taken to be the total number of days from the recruitment date to diabetes diagnosis or the censoring date. This form of administrative censoring implies that Assumption 1 is satisfied. 

As 1 of 26 centres contributing to the EPIC-InterAct case-cohort study \citep{Langenberg11}, a random subcohort of size 1025, along with the remaining 863 incident cases, were selected to have their blood samples analysed for fatty acid decomposition. The quantities of the fatty acids were expressed as a percentage of total plasma phospholipid fatty acids (mol\%). Among the 27 fatty acids with relative concentrations greater than $0.05\%$, 9 were identified as saturated fatty acids (SFAs), belonging to 3 different groups: 2 odd-chain SFAs (pentadecanoic acid, C15:0; heptadecanoic acid, C17:0), 3 even-chain SFAs (myristic acid, C14:0; palmitic acid, C16:0; stearic acid, C18:0) and 4 very-long-chain SFAs (arachidic acid, C20:0; behenic acid, C22:0; tricosanoic acid, C23:0; lignoceric acid, C24:0). %describe fatty acid data, and separation of saturated fatty acids

As potential confounders of the effects of the saturated fatty acids on incident type 2 diabetes, we identified age at recruitment, sex, waist circumference, body mass index and physical activity index. Additionally, we have chosen to incorporate 5 dietary variables from the questionnaires to help predict the missing values of the fatty acids. These are daily intakes (grams per day) of: potatoes and other tubers, fruit, fish and shellfish, meat and meat products, and dairy products. 

Individuals with prevalent type 2 diabetes (855 individuals) or unknown diabetes status (5 individuals), as well as those with missing confounder (1832 individuals) or dietary data (310 individuals), were excluded from analysis. Following \citet{Forouhi14}, we also excluded individuals with a ratio of energy intake to energy requirement in the bottom or top 1\% as probable dietary misreporters (432 individuals). There remain 22219 individuals in the dataset, with a subcohort of size 886 (860 controls and 26 incident cases) and 771 non-subcohort incident cases. From this, 14 subcohort individuals and 95 non-subcohort incident cases have missing fatty acid measurements. Instead of excluding these individuals and losing valuable data on cases, we have chosen to assume that this missingness is independent of the values of the missing fatty acid data given the available information, so that Assumption 2 is still satisfied. 

\subsection{Model specification}

We set $W$ to be the potential confounders described in \S\ref{studes}. Sex was represented by a binary variable. The physical activity index data were categorical with four levels: ``Inactive'', ``Moderately inactive'', ``Moderately active'' and ``Active''. This information was decomposed into three binary dummy variables with ``Active'' as the reference category. The remaining confounders---age, waist circumference, and body mass index---were scaled by their full cohort standard deviations. The auxiliary variable $X$ was set to be the 5 dietary variables after undergoing the log-transformation $x \mapsto \log(1+x)$. 

The fatty acid data are compositional---the relative concentrations of the individual fatty acids sum to 100\%. To address this, we applied the additive logratio transformation \citep{Aitchison82}. Denote a fatty acid measurement value by $z^{\prime} = (z^{\prime}_{1}, \ldots, z^{\prime}_{9}, z^{\prime}_{O})$, where $z^{\prime}_{1}, \ldots, z^{\prime}_{9}$ are the relative concentrations of the 9 SFAs, and $z^{\prime}_{O}$ is the total relative concentration of all remaining fatty acids. If all entries of $z^{\prime}$ are non-zero, its additive logratio image in $\mathbb{R}^{9}$ is
\begin{equation} \label{alr}
    \left(\log \frac{z^{\prime}_{1}}{z^{\prime}_{O}}, \ldots, \log \frac{z^{\prime}_{9}}{z^{\prime}_{O}}\right).
\end{equation}
Otherwise, we first take the zero replacement strategy described in \citet{Greenacre19}. Any zero entries of $z^{\prime}$ are replaced by half of the smallest possible positive measurement. In this case, since measurements are given to two decimal places of a percentage, all zeros are replaced by 0.005\%. Set $Z$ to be the transformed fatty acid vector as described after scaling each component by its standard deviation within the subcohort. In \S\ref{app::anal}, we discuss interpretations and the advantages over direct use of the compositional data.

Let $V = (1, W^{\T}, X^{\T})^{\T}$. We specify a multivariate normal linear regression model
\begin{equation} \label{eqn:reg}
    Z \mid W, X, \xi, \Sigma \sim \mathcal{N}(\xi^{\T}V, \Sigma)
\end{equation}
where $\xi \in \mathbb{R}^{13 \times 9}$ and $\Sigma \in \mathbb{R}^{9 \times 9}$. Let $n_{\mathcal{S}} = |\mathcal{S}| = 1548$, the total number of individuals with fatty acid measurements. We use the Jeffreys prior 
\begin{align*}
p(\xi , \Sigma) \propto |\Sigma|^{-(9+1)/2} =|\Sigma|^{-5},
\end{align*}
which can be interpreted as the noninformative limit of a matrix normal-inverse Wishart prior \citep{Gelman13}. By conjugacy, the restricted posterior distributions are
\begin{align} 
\label{eqn:xi} \xi  \mid \Sigma, Z_{\mathcal{S}}, W_{\mathcal{S}}, X_{\mathcal{S}} &\sim \mathcal{MN}(\hat{\xi}, (V_{\mathcal{S}}^{\T}V_{\mathcal{S}})^{-1}, \Sigma) \\
   \label{eqn:sigma} \Sigma | Z_{\mathcal{S}}, W_{\mathcal{S}}, X_{\mathcal{S}} &\sim \mathcal{IW}(\Psi, n_{\mathcal{S}}),
\end{align}
where $\mathcal{MN}$ and $\mathcal{IW}$ denote the matrix normal and inverse Wishart distributions respectively and
\begin{align*}
    \hat{\xi } &= (V_{\mathcal{S}}^{\T}V_{\mathcal{S}})^{-1}V_{\mathcal{S}}^{\T}Z_{\mathcal{S}} \quad \text{(least squares estimator)}\\
    \Psi &= (Z_{\mathcal{S}}-V_{\mathcal{S}}\hat{\xi })^{\T}(Z_{\mathcal{S}}-V_{\mathcal{S}}\hat{\xi })  \quad \text{(residual sum of squares).}
\end{align*}
The remaining notation follows \S\ref{sec:meth}. For the log-hazard ratio $\beta$, we specified independent, weakly informative Student-$t$ priors for each of the components, all centered at 0 with 3 degrees of freedom. 

\subsection{Results for the EPIC-Norfolk data} \label{app::anal}

For the application, the size and complexity of the dataset necessitated a correlated sampling algorithm to achieve good mixing; we took the approach described at the end of \S\ref{sec:corr}, correlating both the missing fatty acid variables $Z^{\text{mis}}$ and the regression coefficients $\xi$. The full details are provided in \autoref{appen::comp}. We discarded the first 200000 iterations of the sampler, and used the following 800000 for analysis. The convergence diagnostics can be found in \autoref{sec::diag}. 

To interpret the results, we recall that the fatty acid data---originally compositional---were additive logratio transformed using (\ref{alr}), and then scaled by their respective estimated standard deviations. For concreteness, let us specifically consider the saturated fatty acid C14:0. The posterior mean estimate of the hazard ratio is 1.18 (Table \ref{tab:tab5}), implying that an increase of 1 standard deviation in the logratio corresponding to C14:0, keeping all other logratios and confounders fixed, increases the hazard of type 2 diabetes onset by 18\%. Framing this with respect to a particular individual, the change occurs if their \textit{absolute} quantity of C14:0 increases, with all else kept equal. This way, the only logratio that changes is the one corresponding to C14:0; the ratios of the other saturated fatty acids to the reference category (the total of all remaining fatty acids) remain the same as before. Cox regression with isometric logratio transformed compositional data has previously been proposed \citep{McGregor20}, but this produces much less interpretable results than what is described above.
%Previous studies used the relative concentrations of the fatty acids without transformation; a review of these studies and a meta-analysis can be found in \citet{Huang19}.

A review and meta-analysis of previous studies can be found in \citet{Huang19}. To the best of our knowledge, our work is the first to use transformed fatty acid data to investigate this problem. There are several reasons why we believe that this is preferable over direct use of the raw data. First, as noted by \citet{Pearson1897}, treating proportions as absolutely measurements runs the risk of introducing ``spurious correlation'' into the analysis. In \autoref{fig:heatcorr}, we observe that the moderate negative correlation on the original scale between C16:0 and C18:0---by far the two most abundant saturated fatty acids---is removed after transformation. Also, additive changes in percentages ignore the inherently relative nature of the data. For example, an increase from 0\% to 1\% of a fatty acid is viewed as equivalent to an increase from 4\% to 5\%. One could further argue that increasing the proportion of a single fatty acid while keeping some others fixed does not correspond to any type of meaningful hypothetical intervention. Moreover, the total proportion of all omitted fatty acids (e.g. all non-saturated fatty acids, or everything apart from the even-chain SFAs) is forced to decrease in order for the proportions to sum to 100\%, making the analysis strongly dependent on the choice of included fatty acids. This could partly explain the disparity in results across studies. In contrast, our use of the transformation gives us the previously described interpretation of increasing the absolute quantity of a fatty acid. This corresponds to a more intuitive intervention, and only depends on the particular fatty acid that is being changed.

The meta-analysis by \citet{Huang19}---with 10 studies included---suggested that there was conclusive evidence for the effects of only three saturated fatty acids: C15:0 and C17:0 (inverse association with type 2 diabetes), and C14:0 (positive association). In this regard, our results for C17:0 and C14:0 are consistent with the existing literature. It is less clear-cut for C15:0, although there is a weak indication that an inverse association is present.

Even-chain SFAs account for the bulk of the total amount of saturated fatty acids, and they have been linked to an increased risk of type 2 diabetes in several studies \citep[e.g.][]{Forouhi14, Lu18}. Our results for C14:0 and C16:0 support this link, but no evidence of association was found for C18:0. We conjecture that the disparity for C18:0 can be explained by our use of transformed fatty acid data. On the raw data scale, increasing the proportion of C18:0 while keeping the proportions of the other SFAs fixed forces the total proportion of non-saturated fatty acids to decrease. On the transformed scale, this corresponds to an increase in \textit{all} of the logratios. In \autoref{appen:c18}, we provide an informal calculation that shows how the effects from the other logratios could indicate a positive association for C18:0, even when such an association does not exist. Particularly, the relatively small standard deviation of C16:0 on the transformed scale (Table \ref{tab:tab5}) allows its strong positive association to dominate. This suggests that the effects from C18:0 found by previous studies may in fact be mostly due to C16:0 instead.
%explain how the effect is non-negligible because the sd of c18:0 on the raw scale is relatively large?

Comparatively few studies have investigated the association between very-long-chain SFAs and type 2 diabetes. \citet{Forouhi14} analysed data from the EPIC-InterAct Project, which incorporates data from 26 studies from 8 different countries in Europe, including the EPIC-Norfolk dataset analysed here. This analysis suggested that all four of the very-long-chain SFAs examined here are inversely associated with type 2 diabetes. Our findings for C22:0 differ, instead supporting a positive association, matching the conclusions of \citet{Lin18} using data from a Chinese population. On the other hand, our results indicate inverse associations for C20:0 and C24:0; this heterogeneity within an SFA group supports the argument that the effect of each SFA should be studied separately.

\begin{figure}
\begin{center}
\resizebox{\textwidth}{!}{\includegraphics[width=4.5in]{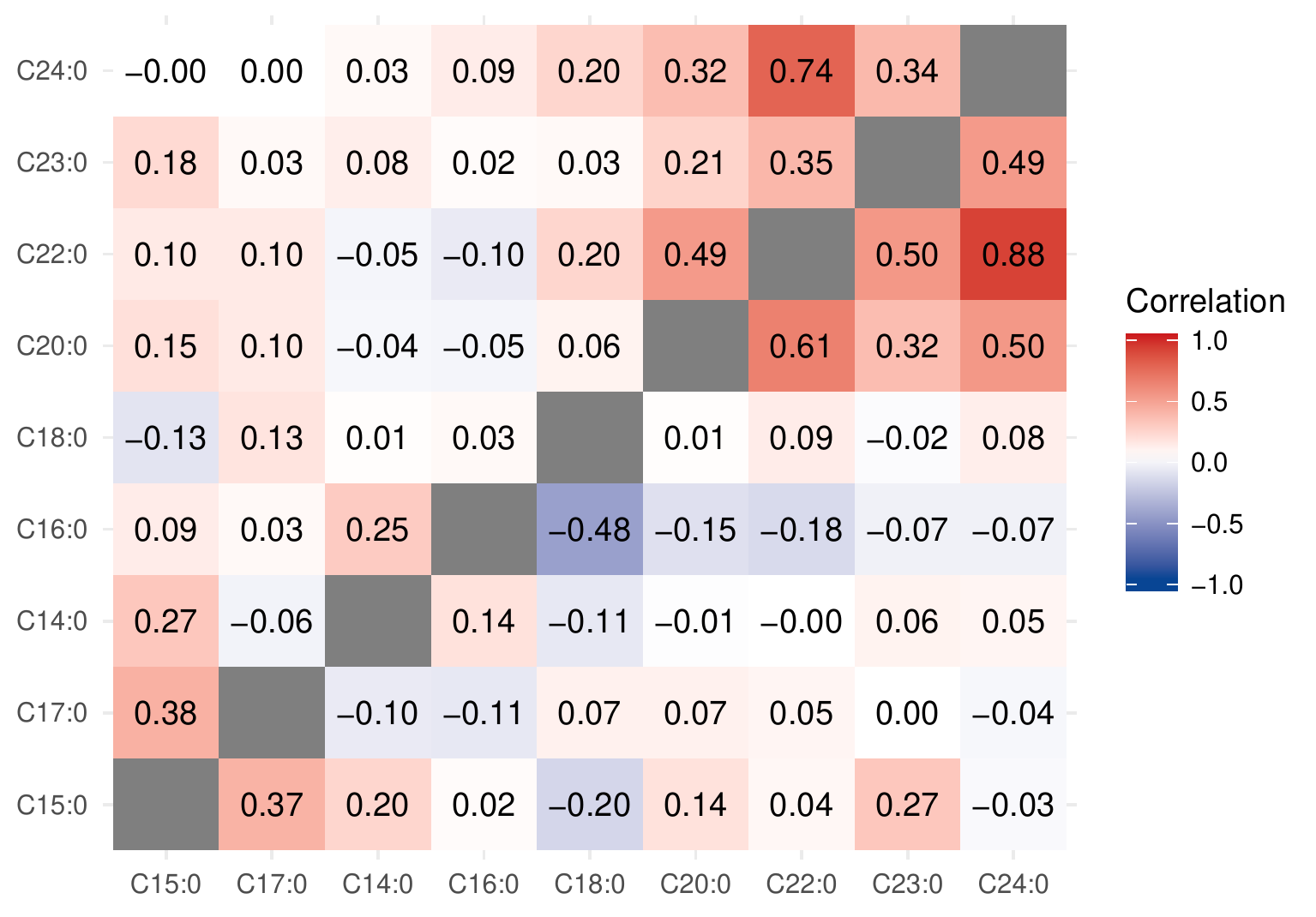}}
\end{center}
\caption{Estimated correlations between the saturated fatty acids using the subcohort data. Values below the diagonal were computed from the raw data; values above the diagonal were computed from the additive logratio transformed data. \label{fig:heatcorr}}
\end{figure}

\begin{table}
\caption{Data summaries for the subcohort individuals with complete data, and analysis results. The raw data are expressed as percentages of the total phospholipid fatty acids. SFA, saturated fatty acid; ocSFAs, odd-chain saturated fatty acids; evSFAs, even-chain saturated fatty acids; vlcSFAs, very-long-chain saturated fatty acids; ALR, additive logratio transformed; SD, standard deviation; HR, hazard ratio. ``HR 95\%'' refers to the central 95\% credible interval; ``$\mathbb{P}(\text{HR} \leq 1)$'' refers to the posterior probability that the hazard ratio does not exceed 1. \label{tab:tab5}}
\begin{center}
\resizebox{\textwidth}{!}{\begin{tabular}{ll|c|c| ccc|}
\hline
\multicolumn{2}{|c|}{} & \multicolumn{1}{c|}{Raw data} &  \multicolumn{1}{c|}{ALR data} & \multicolumn{3}{c|}{Analysis results}\\ \hline
\multicolumn{1}{|l|}{Group} & SFA & Mean (SD)  & Mean (SD) &  HR Mean  & HR 95\% & $\mathbb{P}(\text{HR} \leq 1)$ \\ \hline
\multicolumn{1}{|l|}{ocSFAs}& C15:0 & 0.25\% (0.07\%) & -5.42 (0.27) & 0.97  &  (0.86, 1.10) & 0.684\\
\multicolumn{1}{|l|}{}& C17:0 & 0.43\% (0.09\%) & -4.86 (0.26) &  0.86 &  (0.77, 0.95) & 0.998\\ [2ex]
\multicolumn{1}{|l|}{ecSFAs}& C14:0 & 0.39\% (0.10\%) & -4.95 (0.26) & 1.18  &  (1.05, 1.33) & 0.003\\
\multicolumn{1}{|l|}{}& C16:0 & 30.12\% (1.54\%) & -0.59 (0.07) &  1.39 & (1.24, 1.55) & 0.000\\ 
\multicolumn{1}{|l|}{}& C18:0 & 13.97\%  (1.32\%) & -1.36 (0.11) &  0.99 &  (0.88, 1.12) & 0.585\\ [2ex]
\multicolumn{1}{|l|}{vlcSFAs}& C20:0 & 0.16\% (0.05\%) & -5.89 (0.31) &  0.91 &  (0.82, 1.02) & 0.947\\
\multicolumn{1}{|l|}{}& C22:0 & 0.29\% (0.10\%) & -5.27 (0.24) & 1.11   & (0.96, 1.29) & 0.074 \\
\multicolumn{1}{|l|}{}& C23:0 & 0.14\% (0.07\%) & -6.13 (0.70) & 0.99  & (0.88, 1.11) & 0.601\\ 
\multicolumn{1}{|l|}{}& C24:0 & 0.24\% (0.08\%) & -5.44 (0.26)  &  0.78  & (0.70, 0.87) & 1.000\\
\hline
\end{tabular}}
\end{center}
\end{table}

\begin{figure}
\begin{center}
\resizebox{\textwidth}{!}{\includegraphics[width=4.5in]{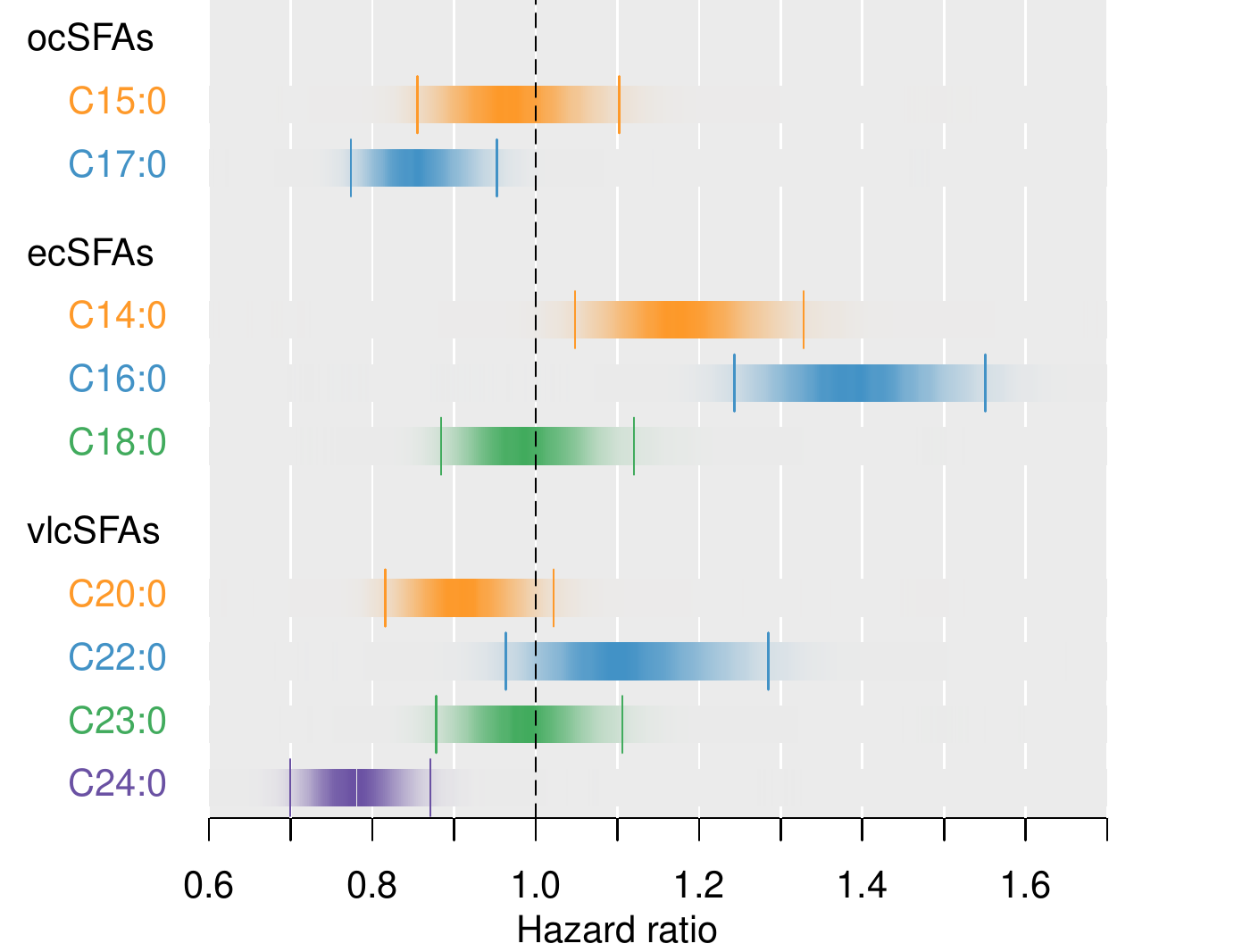}}
\end{center}
\caption{Posterior distributions of the saturated fatty acid hazard ratios. The darkness of the strips is proportional to the posterior density, with the central 95\% credible regions indicated. ocSFAs, odd-chain saturated fatty acids; evSFAs, even-chain saturated fatty acids; vlcSFAs, very-long-chain saturated fatty acids. \label{fig:dens}}
\end{figure}

\section{Discussion}

\iffalse
The method bears a similarity with multiple imputation in that multiple complete datasets are imputed, and a form of averaging across datasets takes place. However, while multiple imputation takes an average of point estimators, our method averages likelihood functions within a fully Bayesian framework. 
\fi

This paper introduces a novel methodology for case-cohort Cox regression. We are able to incorporate auxiliary variables to help predict the missing covariate values and are unrestricted in our choice of prediction model; this differs from multiple imputation \citep{Keogh13}, which requires careful specification of prediction models to avoid incompatability with the Cox model. The models for the nuisance parameters, including the baseline cumulative hazard function, are nonparametrically specified and then integrated out, facilitating robust and convenient inference. By modifying the basic sampling algorithm, the method scales effectively to datasets with a large sample size and a moderate number of covariates, in contrast to nonparametric maximum likelihood estimation \citep{Zeng14}. We demonstrated this scalability in our analysis of the EPIC-Norfolk study. Simulations suggest that we obtain substantial efficiency gains over weighted Cox regression approaches \citep[e.g.][]{Prentice86}, which are the status quo in practice. As part of our analysis of the EPIC-Norfolk study data, we also developed a new approach for handling compositional data in the Cox model that provides more reliable and interpretable results compared to previous studies. 

There is ample scope to extend our framework. We have assumed that the covariates are time-independent since this was sufficient for our application, where only baseline measurements were available. This assumption can be relaxed by building on the results of \citet{Sinha03}, which provided a Bayesian justification of the Cox partial likelihood in various settings.

The nested case-control design \citep{Thomas77} is similar to the case-cohort design in the sense that full covariate measurements are obtained for all cases, but only for a sample of controls. Like the nonparametric maximum likelihood approach of \citet{ScheikeJuul04}, \citet{Scheike04} and \citet{Zeng14}, it is straightforward to adapt our method to the nested case-control design under similar assumptions. Generalizing our method to other survival models like \citet{Zeng14} is an area for future research.

Another important direction for further work is variable selection. Existing proposals are few in number and revolve around weighted Cox regression \citep{Ni16, Newcombe18}. Extending our framework to perform variable selection will not only allow more efficient use of data, but also has the advantage of adopting the principled Bayesian approach to variable selection \citep{Clyde04}. 

\section*{Acknowledgements}

The authors thank Nicola Kerrison (MRC Epidemiology Unit, Cambridge, UK) for managing and providing us with the data from EPIC-Norfolk used in Section 4, and the laboratory team at the MRC Epidemiology Unit for managing the blood samples for the EPIC-InterAct project. Andrew Yiu, Robert J.B. Goudie and Brian D.M. Tom were funded by the UK Medical Research Council programme MRC\_MC\_UU\_00002/2. Stephen J. Sharp was funded by the UK Medical Research Council (MRC; MC\_UU\_12015/1). Paul J. Newcombe was funded by the UK Medical Research Council programme MC\_UU\_00002/9 and also acknowledges support from the NIHR Cambridge BRC. Funding for the EPIC-InterAct project was provided by the EU FP6 Programme (grant number LSHM\_CT\_2006\_037197).

\bibliography{paper-ref}

\begin{thebibliography}{41}
\providecommand{\natexlab}[1]{#1}
\providecommand{\url}[1]{\texttt{#1}}
\expandafter\ifx\csname urlstyle\endcsname\relax
  \providecommand{\doi}[1]{doi: #1}\else
  \providecommand{\doi}{doi: \begingroup \urlstyle{rm}\Url}\fi

\bibitem[Aitchison(1982)]{Aitchison82}
J.~Aitchison.
\newblock The statistical analysis of compositional data.
\newblock \emph{Journal of the Royal Statistical Society, Series B},
  44:\penalty0 139--177, 1982.

\bibitem[Andrieu and Roberts(2009)]{Andrieu09}
C.~Andrieu and G.~Roberts.
\newblock The pseudo-marginal approach for efficient monte carlo computations.
\newblock \emph{Annals of Statistics}, 37:\penalty0 697--725, 2009.

\bibitem[Barlow(1994)]{Barlow94}
W.~E. Barlow.
\newblock Robust variance estimation for the case-cohort design.
\newblock \emph{Biometrics}, 50:\penalty0 1064--1072, 1994.

\bibitem[Borgan and Samuelson(2017)]{Borgan17}
{\O}.~Borgan and S.~Samuelson.
\newblock Cohort sampling for time-to-event data: an overview.
\newblock In {\O}.~Borgan, N.~Breslow, N.~Chatterjee, M.~Gail, A.~Scott, and
  C.~Wild, editors, \emph{Handbook of Statistical Methods for Case-Control
  studies}, pages 285--301. CRC Press, Boca Raton, 2017.

\bibitem[Borgan et~al.(2000)Borgan, Langholz, Samuelson, Goldstein, and
  Pogoda]{Borgan00}
{\O}.~Borgan, B.~Langholz, S.~Samuelson, L.~Goldstein, and J.~Pogoda.
\newblock Exposure stratified case-cohort designs.
\newblock \emph{Lifetime Data Analysis}, 6:\penalty0 39--58, 2000.

\bibitem[Breslow(1972)]{Breslow72}
N.~Breslow.
\newblock Discussion of: {R}egression models and life-tables.
\newblock \emph{Journal of the Royal Statistical Society, Series B},
  34:\penalty0 216--218, 1972.

\bibitem[Chen and Lo(1999)]{Chen99}
K.~Chen and S.~Lo.
\newblock Case-cohort and case-control analysis with {C}ox's model.
\newblock \emph{Biometrika}, 86:\penalty0 755--764, 1999.

\bibitem[Clyde and George(2004)]{Clyde04}
M.~Clyde and E.~George.
\newblock Model uncertainty.
\newblock \emph{Statistical Science}, 19:\penalty0 81--94, 2004.

\bibitem[Cox(1972)]{Cox72}
D.~Cox.
\newblock Regression models and life-tables.
\newblock \emph{Journal of the Royal Statistical Society, Series B},
  34:\penalty0 187--220, 1972.

\bibitem[Deligiannidis et~al.(2018)Deligiannidis, Doucet, and
  Pitt]{Deligiannidis18}
G.~Deligiannidis, A.~Doucet, and M.~Pitt.
\newblock The correlated pseudomarginal method.
\newblock \emph{Journal of the Royal Statistical Society, Series B},
  80:\penalty0 839--870, 2018.

\bibitem[Forouhi et~al.(2014)]{Forouhi14}
N.~Forouhi et~al.
\newblock {Differences in the prospective association between individual plasma
  phospholipid saturated fatty acids and incident type 2 diabetes: the
  EPIC-InterAct case-cohort study}.
\newblock \emph{Lancet Diabetes and Endocrinology}, 2:\penalty0 810--818, 2014.

\bibitem[Gelman et~al.(2013)Gelman, Carlin, Stern, and Rubin]{Gelman13}
A.~Gelman, J.~Carlin, H.~Stern, and A.~Rubin.
\newblock \emph{{Bayesian Data Analysis}}.
\newblock CRC Press, Boca Raton, 2013.

\bibitem[Greenacre(2019)]{Greenacre19}
M.~Greenacre.
\newblock \emph{{{C}ompositional {D}ata {A}nalysis {I}n {P}ractice}}.
\newblock Chapman \& Hall/CRC, New York, 2019.

\bibitem[Huang et~al.(2019)]{Huang19}
L.~Huang et~al.
\newblock Circulating saturated fatty acids and incident type 2 diabetes: A
  systematic review and meta-analysis.
\newblock \emph{Nutrients}, 11:\penalty0 5, 2019.

\bibitem[Kalbfleisch(1978)]{Kalbfleisch78}
J.~Kalbfleisch.
\newblock Non-parametric {B}ayesian analysis of survival time data.
\newblock \emph{Journal of the Royal Statistical Society, Series B},
  40:\penalty0 214--221, 1978.

\bibitem[Kalbfleisch and Lawless(1988)]{Kalbfleisch88}
J.~Kalbfleisch and J.~Lawless.
\newblock Likelihood analysis for multi-state models for disease incidence and
  mortality.
\newblock \emph{Statistics in Medicine}, 7:\penalty0 149--160, 1988.

\bibitem[Keogh and White(2013)]{Keogh13}
R.~Keogh and I.~White.
\newblock Using full-cohort data in nested case-control and case-cohort studies
  by multiple imputation.
\newblock \emph{Statistics in Medicine}, 32:\penalty0 4021--4043, 2013.

\bibitem[Kim and Lee(2003)]{Kim03}
Y.~Kim and J.~Lee.
\newblock Bayesian bootstrap for proportional hazards model.
\newblock \emph{Annals of Statistics}, 31:\penalty0 1905--1922, 2003.

\bibitem[Kulathinal and Arjas(2006)]{Kulathinal06}
S.~Kulathinal and E.~Arjas.
\newblock Bayesian inference from case-cohort data with multiple end-points.
\newblock \emph{Scandinavian Journal of Statistics}, 33:\penalty0 25--36, 2006.

\bibitem[Kulich and Lin(2004)]{Kulich04}
M.~Kulich and D.~Y. Lin.
\newblock Improving the efficiency of relative-risk estimation in case-cohort
  studies.
\newblock \emph{Journal of the American Statistical Association}, 99:\penalty0
  832--844, 2004.

\bibitem[Langenberg et~al.(2011)]{Langenberg11}
C.~Langenberg et~al.
\newblock {Design and cohort description of the InterAct Project: an
  examination of the interaction of genetic and lifestyle factors on the
  incidence of type 2 diabetes in the EPIC Study}.
\newblock \emph{Diabetologia}, 54:\penalty0 2272--2282, 2011.

\bibitem[Lin(2018)]{Lin18}
J.~Lin.
\newblock Erythrocyte saturated fatty acids and incident type 2 diabetes in
  {C}hinese men and women: A prospective cohort study.
\newblock \emph{Nutrients}, 10:\penalty0 1393, 2018.

\bibitem[Lu and Tsiatis(2006)]{Lu06}
W.~Lu and A.~Tsiatis.
\newblock Semiparametric transformation models for the case-cohort study.
\newblock \emph{Biometrika}, 93:\penalty0 207--214, 2006.

\bibitem[Lu et~al.(2018)]{Lu18}
Y.~Lu et~al.
\newblock Serum lipids in association with type 2 diabetes risk and prevalence
  in a {C}hinese population.
\newblock \emph{Journal of Clinical Endocrinology and Metabolism},
  103:\penalty0 671--680, 2018.

\bibitem[McGregor et~al.(2020)McGregor, Palarea-Albaladejo, Dall, Hron, and
  Chastin]{McGregor20}
D.~McGregor, J.~Palarea-Albaladejo, P.~Dall, K.~Hron, and S.~Chastin.
\newblock Cox regression survival analysis with compositional covariates:
  Application to modelling mortality risk from 24-h physical activity patterns.
\newblock \emph{Statistical Methods in Medical Research}, 29:\penalty0
  1386--1402, 2020.

\bibitem[Nan et~al.(2004)Nan, Emond, and Wellner]{Nan04a}
B.~Nan, M.~Emond, and J.~Wellner.
\newblock {Information bounds for Cox regression models with missing data}.
\newblock \emph{Annals of Statistics}, 32:\penalty0 723--753, 2004.

\bibitem[Newcombe et~al.(2018)Newcombe, Connolly, Seaman, Richardson, and
  Sharp]{Newcombe18}
P.~Newcombe, S.~Connolly, S.~Seaman, S.~Richardson, and S.~Sharp.
\newblock A two-step method for variable selection in the analysis of a
  case-cohort study.
\newblock \emph{International Journal of Epidemiology}, 47:\penalty0 597--604,
  2018.

\bibitem[Ni et~al.(2016)Ni, Cai, and Zeng]{Ni16}
A.~Ni, J.~Cai, and D.~Zeng.
\newblock Variable selection for case-cohort studies with failure time outcome.
\newblock \emph{Biometrika}, 103:\penalty0 547--562, 2016.

\bibitem[Pearson(1897)]{Pearson1897}
K.~Pearson.
\newblock Mathematical contributions to the theory of evolution on a form of
  spurious correlation which may arise when indices are used in the measurement
  of organs.
\newblock \emph{Proceedings of the Royal Society of London LX}, pages 489--502,
  1897.

\bibitem[Prentice(1986)]{Prentice86}
R.~Prentice.
\newblock A case-cohort design for epidemiologic cohort studies and disease
  prevention trials.
\newblock \emph{Biometrika}, 73:\penalty0 1--11, 1986.

\bibitem[Rubin(1981)]{Rubin81}
D.~Rubin.
\newblock The {B}ayesian bootstrap.
\newblock \emph{Annals of Statistics}, 9:\penalty0 130--134, 1981.

\bibitem[Scheike and Juul(2004)]{ScheikeJuul04}
T.~Scheike and A.~Juul.
\newblock {Maximum likelihood estimation for Cox's regression model under
  nested case-control sampling}.
\newblock \emph{Biostatistics}, 5:\penalty0 193--206, 2004.

\bibitem[Scheike and Martinussen(2004)]{Scheike04}
T.~Scheike and T.~Martinussen.
\newblock Maximum likelihood estimation for {C}ox's regression model under
  case-cohort sampling.
\newblock \emph{Scandinavian Journal of Statistics}, 31:\penalty0 283--293,
  2004.

\bibitem[Sharp et~al.(2014)Sharp, Poulaliou, Thompson, White, and
  Wood]{Sharp14}
S.~Sharp, M.~Poulaliou, S.~Thompson, I.~White, and A.~Wood.
\newblock A review of published analyses of case-cohort studies and
  recommendations for future reporting.
\newblock \emph{PLoS One}, 9:\penalty0 e101176, 2014.

\bibitem[Sherlock et~al.(2017)Sherlock, Thiery, and Lee]{Sherlock17}
C.~Sherlock, A.~Thiery, and A.~Lee.
\newblock Pseudo-marginal {M}etropolis-{H}astings sampling using averages of
  unbiased estimators.
\newblock \emph{Biometrika}, 104:\penalty0 727--734, 2017.

\bibitem[Sinha et~al.(2003)Sinha, Ibrahim, and Chen]{Sinha03}
D.~Sinha, J.~Ibrahim, and M.~Chen.
\newblock A {B}ayesian justification of {C}ox's partial likelihood.
\newblock \emph{Biometrika}, 90:\penalty0 629--641, 2003.

\bibitem[Steingrimsson and Strawderman(2017)]{Steingrimsson17}
J.~Steingrimsson and R.~Strawderman.
\newblock Estimation in the semiparametric accelerated failure time model with
  missing covariates: Improving efficiency through augmentation.
\newblock \emph{Journal of the American Statistical Association}, 112:\penalty0
  1221--1235, 2017.

\bibitem[Tanner and Wong(1987)]{Tanner87}
M.~Tanner and W.~Wong.
\newblock The calculation of posterior distributions by data augmentation.
\newblock \emph{Journal of the American Statistical Association}, 82:\penalty0
  528--540, 1987.

\bibitem[Thomas(1977)]{Thomas77}
D.~Thomas.
\newblock Addendum to: ``{M}ethods of cohort analysis: appraisal by application
  to asbestos mining,'' by {L}iddell, {F}.{D}.{K}., {M}c{D}onald, {J}.{C}. and
  {T}homas, {D}.{C}.
\newblock \emph{Journal of the Royal Statistical Society, Series A},
  140:\penalty0 469--491, 1977.

\bibitem[van~der Vaart(1998)]{vanderVaart98}
A.~van~der Vaart.
\newblock \emph{{Asymptotic Statistics}}.
\newblock Cambridge University Press, Cambridge, 1998.

\bibitem[Zeng and Lin(2014)]{Zeng14}
D.~Zeng and D.~Lin.
\newblock Efficient estimation of semiparametric transformation models for
  two-phase cohort studies.
\newblock \emph{Journal of the American Statistical Association}, 109:\penalty0
  371--383, 2014.

\end{thebibliography}
\newpage
\appendix
\section{Derivation of the marginal posterior of $\beta$} \label{appen::deriv}
We provide a more detailed derivation of expression (\ref{eqn:margpost}). First, (\ref{eqn:post}) is proportional to
\begin{equation*}
\begin{split}
    \left[ \prod_{i\in \mathcal{S}} \exp(\beta_{1}^{\T}Z_{i}+\beta_{2}^{\T}W_{i})^{\Delta_{i}}\exp\left\{-e^{\beta_{1}^{\T}Z_{i}+\beta_{2}^{\T}W_{i}}\Lambda_{0}(Y_{i})\right\}\right]\\ \left[\prod_{j\in \bar{\mathcal{S}}}\int\exp\left\{-e^{\beta_{1}^{\T}z_{j}+\beta_{2}^{\T}W_{j}}\Lambda_{0}(Y_{j})\right\}p(z_{j} \mid W_{j},X_{j}, \gamma) dz_{j} \right] p(\gamma \mid D_{\mathcal{S}})p(\beta),
    \end{split}
\end{equation*}
where we have incorporated the restricted posterior of $\gamma$. Then, we integrate with respect to $\Lambda_{0}$ and apply Fubini's theorem to bring the $\Lambda_{0}$ integral inside:
\begin{equation}
\label{eqn:apppost}
\begin{split}
    \int_{\{z_{j}: j \in \bar{\mathcal{S}}\}} \int_{\Lambda_{0}}\left[ \prod_{i\in \mathcal{S}} \exp(\beta_{1}^{\T}Z_{i}+\beta_{2}^{\T}W_{i})^{\Delta_{i}}\exp\left\{-e^{\beta_{1}^{\T}Z_{i}+\beta_{2}^{\T}W_{i}}\Lambda_{0}(Y_{i})\right\}\right]\\ \left[\prod_{j\in \bar{\mathcal{S}}}\exp\left\{-e^{\beta_{1}^{\T}z_{j}+\beta_{2}^{\T}W_{j}}\Lambda_{0}(Y_{j})\right\}\right]d\Lambda_{0}\left[\prod_{k\in \bar{\mathcal{S}}}p(z_{k} \mid W_{k},X_{k}, \gamma) dz_{k} \right]  p(\gamma \mid D_{\mathcal{S}})p(\beta).
    \end{split}
\end{equation}
The $\Lambda_{0}$ integral on the inside can be rewritten as
\begin{equation*}
\begin{split}
     \int_{\Lambda_{0}} \prod_{k =1}^{n} \left[ \exp(\beta_{1}^{\T}\tilde{Z}_{k}+\beta_{2}^{\T}W_{k})  \exp\left\{-\Delta\Lambda_{0}(Y_{k}) \sum_{l=1}^{n} R_{l}(T_{k})e^{\beta_{1}^{\T}\tilde{Z}_{l}+\beta_{2}^{\T}W_{l}}\right\}\right]^{\Delta_{k}}d\Lambda_{0}
    \end{split}
\end{equation*}
where $\tilde{Z}_{k}$ equals $Z_{k}$ if $k \in \mathcal{S}$ and equals $z_{k}$ otherwise. Integrating out each  $\Delta \Lambda_{0}(Y_{k})$ yields
\begin{equation*} 
    \prod_{k = 1}^n \left\{\frac{\exp{(\beta_{1}^{\T}\tilde{Z}_{k}}+\beta_{2}^{\T}W_{k})}{\sum_{l=1}^n R_{l}(T_{k})\exp{(\beta_{1}^{\T}\tilde{Z}_{l}+\beta_{2}^{\T}W_{l})} }\right\}^{\Delta_{k}}.
\end{equation*}
Substituting this back into (\ref{eqn:apppost}) and then integrating with respect to $\gamma$ yields (\ref{eqn:margpost}).

\section{Justification of Algorithm 2} \label{appen:alg2}

Let $\boldsymbol{\gamma} = (\gamma_{1}, \ldots, \gamma_{B})$. To justify Algorithm \ref{al2}, it is sufficient to check that detailed balance holds for $(U, \boldsymbol{\gamma})$. This amounts to showing that
\begin{equation} \label{dbal}
    \phi(u; 0_{M}, I_{M})p(\boldsymbol{\gamma} \mid \{D_{i}: i \in \mathcal{S}\})K\{(u, \boldsymbol{\gamma}), (\Tilde{u}, \Tilde{\boldsymbol{\gamma}})\} =  \phi(\Tilde{u}; 0_{M}, I_{M})p(\Tilde{\boldsymbol{\gamma}} \mid \{D_{i}: i \in \mathcal{S}\})K\{(\Tilde{u}, \Tilde{\boldsymbol{\gamma}}), (u, \boldsymbol{\gamma})\}
\end{equation}
where $\phi(\cdot; \mu, \Sigma)$ is the density function of $\mathcal{N}(\mu, \Sigma)$ and $K\{(u, \boldsymbol{\gamma}), (\Tilde{u}, \Tilde{\boldsymbol{\gamma}})\} = \phi(\Tilde{u}; \rho u, (1-\rho^{2})I_{M}) p(\Tilde{\boldsymbol{\gamma}} \mid \{D_{i}: i \in \mathcal{S}\})$. Clearly, the terms involving $\boldsymbol{\gamma}$ on both sides of (\ref{dbal}) match. Furthermore,
\begin{align*}
    \phi(u; 0_{M}, I_{M})\phi(\Tilde{u}; \rho u, (1-\rho^{2})I_{M}) &= (2\pi)^{-M}(1-\rho^{2})^{-M/2}\exp{\left\{\frac{1}{2}\left[u^{\T}u+\frac{(\Tilde{u}-\rho u)^{\T}(\Tilde{u}-\rho u)}{1-\rho^{2}}\right]\right\}} \\
    &= (2\pi)^{-M}(1-\rho^{2})^{-M/2}\exp{\left\{\frac{1}{2}\left[\Tilde{u}^{\T}\Tilde{u}+\frac{(u-\rho \Tilde{u})^{\T}(u-\rho \Tilde{u})}{1-\rho^{2}}\right]\right\}} \\
    &= \phi(\Tilde{u}; 0_{M}, I_{M})\phi(u; \rho \Tilde{u}, (1-\rho^{2})I_{M}),
\end{align*}
which establishes (\ref{dbal}).

\section{Application computation} \label{appen::comp}

We set $B=1$. First, consider sampling $\xi$ given $(\Sigma, Z_{\mathcal{S}}, W_{\mathcal{S}}, X_{\mathcal{S}})$. Let $C = (V_{\mathcal{S}}^{\T}V_{\mathcal{S}})^{-1}$. Since $C$ and $\Sigma$ are both positive definite, they possess unique positive definite square roots $C^{1/2}$ and $\Sigma^{1/2}$ respectively. Let $U_{\xi} \sim \mathcal{MN}(0_{13\times9}, I_{13\times13}, I_{9 \times 9})$---or equivalently, let $U_{\xi}$ be a $13\times 9$ matrix where the entries are independent $\mathcal{N}(0,1)$ variables---independent of $(\Sigma, Z_{\mathcal{S}}, W_{\mathcal{S}}, X_{\mathcal{S}})$. Then,
\begin{equation*}
    \varphi_{\xi}(U_{\xi}, \Sigma, Z_{\mathcal{S}}, W_{\mathcal{S}}, X_{\mathcal{S}}) = \hat{\xi} + C^{1/2}U_{\xi}\Sigma^{1/2}
\end{equation*}
has the conditional distribution (\ref{eqn:xi}).

Next, consider sampling $Z^{\text{mis}}$ given $(W_{\bar{\mathcal{S}}}, X_{\bar{\mathcal{S}}}, \xi, \Sigma)$. With $U_{Z} \sim \mathcal{N}(0_9, I_{9 \times 9})$ independent of $(W_{\bar{\mathcal{S}}}, X_{\bar{\mathcal{S}}}, \xi, \Sigma)$,
\begin{equation*}
    \varphi_{Z}(U_{Z}, W_{\bar{\mathcal{S}}}, X_{\bar{\mathcal{S}}}, \xi, \Sigma) = \Sigma^{1/2}U_{Z} + \xi^{\T}V_{\bar{\mathcal{S}}}
\end{equation*}
has conditional distribution equal to (\ref{eqn:reg}) for the missing values of $Z$.

The sampling algorithm is described in Algorithm \ref{al3}. The correlation parameters $\rho_{\xi}$ and $\rho_{Z}$ were both set to 0.995. For both the synthetic data experiment and the real application dataset, we used a normal proposal for $\beta$: $q(\cdot \mid \beta) = \mathcal{N}(\beta, V_{\text{prop}})$. Our initial parameter values $\beta^{(0)}$ and proposal variances $V_{\text{prop}}$ are provided in the supplementary code.

\begin{algorithm}[!h]
\caption{Correlated sampling algorithm for the application} \label{al3}
\begin{tabbing}
   \enspace Select an initial parameter value $\beta^{(0)}$\\
   \enspace Draw an initial value $(U^{(0)}_{\xi},U^{(0)}_{Z}, \Sigma^{(0)})$.\\
   \enspace Compute $\xi^{(0)}=\varphi_{\xi}(U_{\xi}^{(0)}, \Sigma^{(0)}, Z_{\mathcal{S}}, W_{\mathcal{S}}, X_{\mathcal{S}})$. \\
   \enspace Compute $Z^{\text{mis}}_{(0)}=\varphi_{Z}(U_{Z}^{(0)}, W_{\bar{\mathcal{S}}}, X_{\bar{\mathcal{S}}}, \xi^{(0)}, \Sigma^{(0)})$. \\
   \enspace For $r=1$ to $r = N$ \\
   \qquad (a) Propose $\Tilde{\beta}$ from $q(\beta \mid \beta^{(r-1)})$.\\
   \qquad (b) Propose $\Tilde{\Sigma}$ from (\ref{eqn:sigma}).\\
   \qquad (c) Sample $\varepsilon_{\xi} \sim  \mathcal{MN}(0_{13\times9}, I_{13\times13}, I_{9 \times 9})$ and set $\Tilde{U}_{\xi} = \rho_{\xi} U^{(r-1)}_{\xi} + \sqrt{(1-\rho_{\xi}^{2})} \varepsilon_{\xi}$ \\
   \qquad (d) Compute $\Tilde{\xi} = \varphi_{\xi}(\Tilde{U}_{\xi}, \Tilde{\Sigma}, Z_{\mathcal{S}}, W_{\mathcal{S}}, X_{\mathcal{S}})$.\\
   \qquad (e) Sample $\varepsilon_{Z} \sim  \mathcal{N}(0_9, I_{9 \times 9})$ and set $\Tilde{U}_{Z} = \rho_{Z} U^{(r-1)}_{Z} + \sqrt{(1-\rho_{Z}^{2})} \varepsilon_{Z}$ \\
   \qquad (f) Compute $\Tilde{Z}^{\text{mis}} = \varphi_{Z}(\Tilde{U}_{Z}, W_{\bar{\mathcal{S}}}, X_{\bar{\mathcal{S}}}, \Tilde{\xi}, \Tilde{\Sigma})$.\\
   \qquad (g) With probability  $\min\left\{1, \frac{q( \beta^{(r-1)} \mid \Tilde{\beta}) p(\Tilde{\beta})h(\Tilde{\beta}, \Tilde{Z}^{\text{mis}})}{q(\Tilde{\beta} \mid \beta^{(r-1)})p(\beta^{(r-1)}) h(\beta^{(r-1)}, Z_{(r-1)}^{\text{mis}})} \right\}$, \\ \qquad set $(\beta^{(r)}, U^{(r)}_{\xi}, U^{(r)}_{Z}) = (\Tilde{\beta}, \Tilde{U}_{\xi}, \Tilde{U}_{Z})$. \\
    \qquad Otherwise, set $(\beta^{(r)}, U^{(r)}_{\xi}, U^{(r)}_{Z}) =(\beta^{(r-1)}, U^{(r-1)}_{\xi}, U^{(r-1)}_{Z})$.\\
    \enspace Output $(\beta^{(1)}, \ldots, \beta^{(N)})$.
\end{tabbing}
\vspace*{-6pt}
\end{algorithm}

\section{Convergence diagnostics} \label{sec::diag}

We provide convergence diagnostics for the sampling computation in \S\ref{app::anal}. \autoref{fig:trace} contains the trace plots for the log-hazard ratios of the nine saturated fatty acids for 3 separate chains, each run for 1000000 iterations.

In \S\ref{app::anal}, we discarded the first 200000 iterations of the sampler and used the subsequent 800000 iterations for analysis. Using the final 800000 iterations for each of the 3 chains, we computed the Gelman-Rubin statistics \citep{Gelman13} for the log-hazard ratios of the 9 saturated fatty acids to be: 1.000019, 1.000053, 1.000005, 1.000056, 1.000052, 1.000021, 1.000082, 1.000057, 1.000033 for C15:0, C17:0, C14:0, C16:0, C18:0, C20:0, C22:0, C23:0 and C24:0 respectively.

\begin{figure}
\begin{center}
\resizebox{\textwidth}{!}{\includegraphics[width=4.5in]{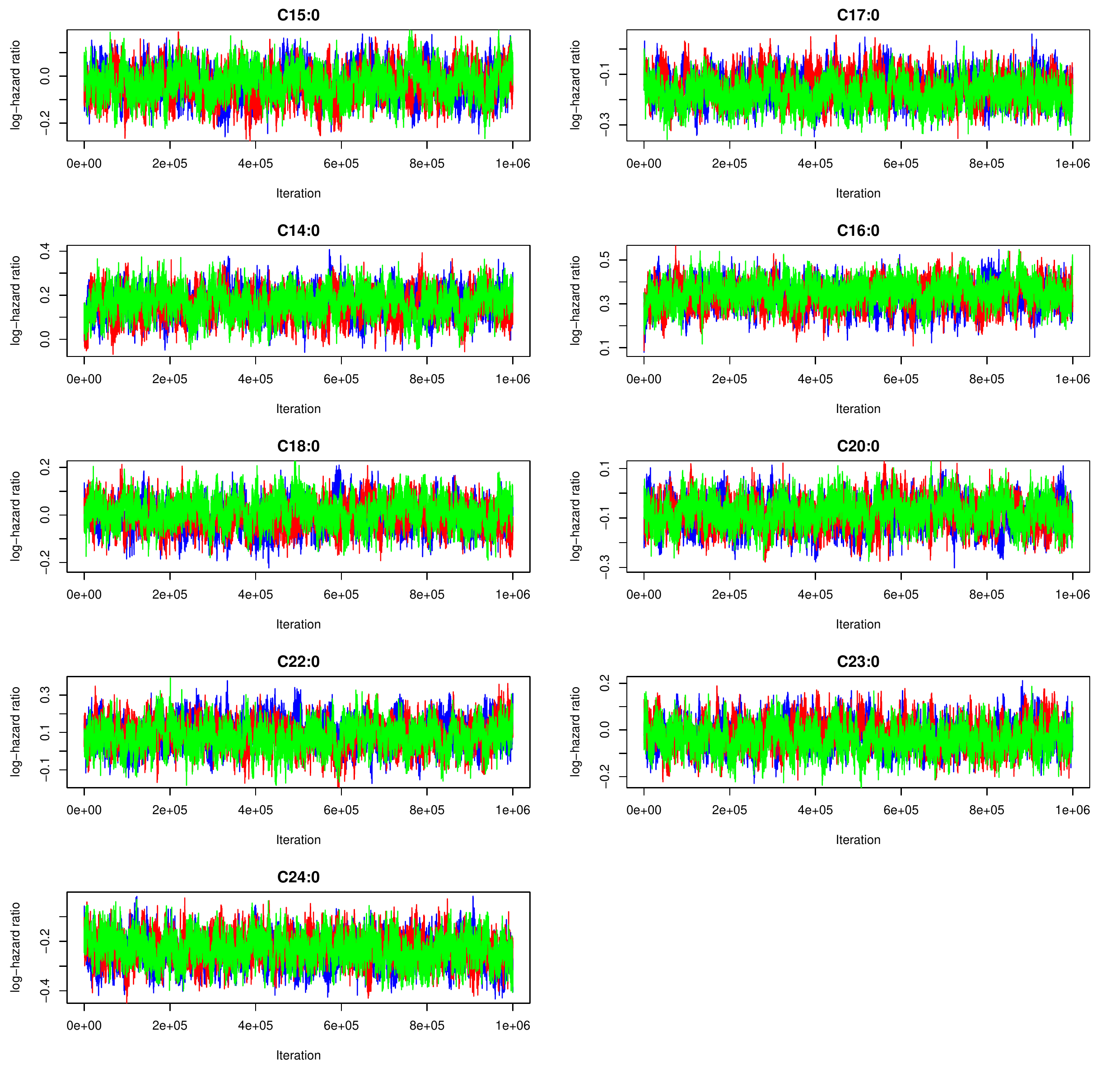}}
\end{center}
\caption{Trace plots for the log-hazard ratios of the nine saturated fatty acids.\label{fig:trace}}

\end{figure}

\section{Investigating the results for C18:0} \label{appen:c18}

In this section, we provide an informal calculation to demonstrate how increasing the relative concentration of C18:0 could indicate a positive association with type 2 diabetes, even when one does not exist on the transformed scale. 

Suppose that our initial saturated fatty acid proportions are equal to the mean values in \autoref{tab:tab5}. This implies that the initial proportion of non-saturated fatty acids is 54.01\%. If we increase the proportion of the fatty acid C18:0 by 1 standard deviation---1.32\%---while keeping the other saturated fatty acid proportions fixed, the proportion of non-saturated fatty acids decreases to 52.69\%. As a result, all logratios apart from the one corresponding to C18:0 increase by $\log(54.01)-\log(52.69) = 0.025$ (3 decimal places). Setting the posterior mean estimates of the hazard ratios in \autoref{tab:tab5} as the truths, we can compute the change in risk as follows:
\begin{align*}
    (0.97)^{\frac{0.025}{0.27}} \cdot (0.86)^{\frac{0.025}{0.26}}\cdot  (1.18)^{\frac{0.025}{0.26}}\cdot  (1.39)^{\frac{0.025}{0.07}} \cdot  (0.91)^{\frac{0.025}{0.31}}\cdot  (1.11)^{\frac{0.025}{0.24}}\cdot  (0.99)^{\frac{0.025}{0.70}}\cdot  (0.78)^{\frac{0.025}{0.26}} \\
    = 1.00 \cdot 0.99 \cdot 1.02 \cdot 1.12 \cdot 0.99 \cdot 1.01 \cdot 1.00 \cdot 0.98\\
    = 1.10.
\end{align*}
We observe in particular that the effect is dominated by the factor of 1.12 from C16:0 due to its small standard deviation (0.07) on the transformed scale. For reference, \citet{Forouhi14} estimated the hazard ratio of C18:0 across 6 different models to be (1.25, 1.06, 1.06, 1.12, 1.12, 1.07).

}
\end{document}